\documentclass[conference]{IEEEtran}
\IEEEoverridecommandlockouts
\usepackage{url}
\usepackage{hyperref}
\usepackage{cite}
\usepackage{amsmath,amssymb,amsfonts}
\usepackage{algorithmic}
\usepackage{graphicx}
\usepackage{textcomp}
\usepackage{xcolor}
\usepackage{makecell}
\usepackage{multirow}
\usepackage{etoolbox}

\def\BibTeX{{\rm B\kern-.05em{\sc i\kern-.025em b}\kern-.08em
    T\kern-.1667em\lower.7ex\hbox{E}\kern-.125emX}}

\makeatletter
\patchcmd{\@maketitle}
  {\addvspace{0.5\baselineskip}\egroup}
  {\addvspace{-1.3\baselineskip}\egroup}
  {}
  {}
\makeatother

\begin{document}

\title{Every Character Counts: From Vulnerability to Defense in Phishing Detection}


\author{\IEEEauthorblockN{Maria Chiper}
\IEEEauthorblockA{
\textit{University of Bucharest}, Romania \\
{\tt chiperm22@gmail.com}}
\and
\IEEEauthorblockN{Radu Tudor Ionescu}
\IEEEauthorblockA{
\textit{University of Bucharest}, Romania \\
{\tt raducu.ionescu@gmail.com}}
\vspace{-1cm}
}

\maketitle

\begin{abstract}
Phishing attacks targeting both organizations and individuals are becoming an increasingly significant threat as technology advances. Current automatic detection methods often lack explainability and robustness in detecting new phishing attacks. In this work, we investigate the effectiveness of character-level deep learning models for phishing detection, which can provide both robustness and interpretability. We evaluate three neural architectures adapted to operate at the character level, namely CharCNN, CharGRU, and CharBiLSTM, on a custom-built email dataset, which combines data from multiple sources. Their performance is analyzed under three scenarios: (i) standard training and testing, (ii) standard training and testing under adversarial attacks, and (iii) training and testing with adversarial examples. Aiming to develop a tool that operates as a browser extension, we test all models under limited computational resources. In this constrained setup, CharGRU proves to be the best-performing model across all scenarios. All models show vulnerability to adversarial attacks, but adversarial training substantially improves their robustness. In addition, by adapting the Gradient-weighted Class Activation Mapping (Grad-CAM) technique to character-level inputs, we are able to visualize which parts of each email influence the decision of each model. Our open-source code and data is released at \url{https://github.com/chipermaria/every-character-counts}.
\end{abstract}

\begin{IEEEkeywords}
phishing detection, adversarial attacks, explainable AI, character-level neural network, adversarial training.
\end{IEEEkeywords}

\setlength{\abovedisplayskip}{2.0pt}
\setlength{\belowdisplayskip}{2.0pt}

\section{Introduction}

 Cyberattacks have become an increasingly pressing issue for society \cite{2019}, as new and more sophisticated attack methods continue to emerge every year, ranging from deceptive emails \cite{paul2024phishingemaildetectionusing} and SMS messages \cite{timko2024quantitativestudysmsphishing} to malicious websites \cite{kalaharsha2021detectingphishingsites} and scripts \cite{jelodar2025largelanguagemodelllm}. As digitization advances at a rapid pace, phishing attempts are also growing in both number and complexity. This evolution renders traditional detection techniques insufficient, as they struggle to keep up with the diversity and ingenuity of modern cyber threats \cite{lim2025explicateenhancingphishingdetection}.

Phishing is among the most prevalent forms of cyberattacks \cite{zia2025webphishingnetwpn}, where attackers aim to deceive users into disclosing confidential information, such as banking credentials or login data. In response to this challenge, the cybersecurity field has turned towards machine learning (ML) as a promising solution for early detection \cite{Opara_2020, Salahdine_2021, article12}. 
ML algorithms have proven significantly more effective than traditional rule-based systems \cite{kuikel2025evaluatinglargelanguagemodels}, as they are often capable of identifying both known threats and novel, sophisticated attacks by recognizing hidden patterns and anomalies within the data. Despite these advantages, machine learning also presents limitations \cite{malik2020hierarchylimitationsmachinelearning}. One of the main concerns is the lack of interpretability of model decisions \cite{fajar2024comparativeanalysisblackboxwhitebox}, an essential requirement in the context of security. Without explainability methods, it is difficult to trust and validate the outcomes of a model. Furthermore, ML models are vulnerable to adversarial attacks \cite{ahmed2024comprehensivereviewadversarialattacks}, i.e.~crafted inputs designed to deceive models into making incorrect predictions. For instance, a manipulated phishing email could bypass detection by being incorrectly labeled as legitimate.

In this research, we address these challenges through several key contributions. First, in order to test detection models in a comprehensive setup, we combine multiple available datasets into a large-scale corpus for email phishing detection. Second, we train and test three efficient deep learning architecture, namely a character-level convolutional neural network (CharCNN) \cite{zhang2015character}, and two character-level recurrent neural networks (RNN), one based on Gated Recurrent Units (CharGRU), and one based on Bidirectional Long Short-Term Memory (CharBiLSTM). We incorporate explainability by integrating a character-level adaptation of the Gradient-weighted Class Activation Mapping (Grad-CAM) technique \cite{selvaraju2017grad} to highlight the most influential character groups of an email contributing to the decision taken by a model. Furthermore, we assess the detection performance of these models against adversarial threats by simulating attacks based on the DeepWordBug method \cite{gao2018black}. Finally, we perform adversarial training to enhance the robustness of the considered models.

In summary, our contribution is twofold:
\begin{itemize}
    \item We aggregate and release a large-scale email dataset for phishing detection.
    \item We adapt Grad-CAM for character-level neural networks and integrate it with various architectures to achieve explainability. 
\end{itemize}

\section{Related Work}

Phishing attacks can take many different forms, as detailed in \cite{inproceedings7}. 
Common types include spear phishing, whaling (targeting high-profile individuals), clone phishing, and business email compromise. Most of these attacks are performed via email and have been a major focus of security researchers over the past years \cite{hazell2023spearphishinglargelanguage, wassermann2023targetedattacksredefiningspear}.
Consequently, the phishing detection problem has been addressed using various approaches \cite{article2,wood2022systematic}. In an early study, Zhang et al.~\cite{article2} analyzed the effectiveness of sender and URL blacklists. They found that these techniques achieved up to 23\% accuracy at the initial time of attack (time zero). Wood et al.~\cite{wood2022systematic} also highlighted the limitations of blacklist-based detection, noting that attackers can quickly switch to new websites or email addresses, making static blacklists easy to evade.
More recently, machine learning techniques have been explored with much better results. For example, in \cite{10.1007/978-3-642-33167-1_47}, the authors processed email content using text pre-processing techniques such as lexical analysis, word normalization (to lowercase), stemming, and stop word removal. Subsequently, by applying multiple ML algorithms to the processed data, they achieved detection accuracies of up to 98.8\%. Another study \cite{altwaijry2024phishing} demonstrated the effectiveness of one-dimensional CNN-based models by reaching performance levels higher than 99\% on popular datasets, such as the Phishing Corpus \cite{listik_phishing_corpus} and the Spam Assassin Dataset \cite{spamassassin2005}. Support Vector Machines (SVM) were also tested in email phishing detection \cite{form2015phishing}, achieving accuracy rates up to 97.25\% on the same datasets. 

Deep-learning algorithms have demonstrated high performance in email phishing detection, employing both textual and metadata features. Vinayakumar et al.~\cite{ra2018deepanti} proposed a model based on n-grams and a Multilayer Perceptron (MLP) on the IWSPA-AP 2018 dataset \cite{fette2018email}, achieving an accuracy of 97.9\%. In \cite{fang2019phishing}, a Recurrent Convolutional Neural Network (RCNN) model evaluated on the Enron \cite{enron} and SpamAssassin \cite{spamassassin} datasets reached exceptional performance levels, higher than 99\%.
Some adaptations of CNN and RNN architectures have been developed to operate at the character level. In \cite{7952603}, the authors presented ROC curves comparing a character-level CNN and the best-performing models on a phishing email dataset, alongside LSTM and GRU architectures. The study of  Aljofey et al.~\cite{shilpa2023phishingcnn} also supported the use of character-level models for phishing classification. 
Another relevant study \cite{inbook5} demonstrated that character-level approaches perform well in combination with RNNs. Specifically, a CharBiGRU-Attention model achieved an accuracy of 99.55\% and an F1-score of 99.54\% in phishing detection. Further contributions include various deep learning architectures. Doshi et al.~\cite{doshi2023comprehensive} combined MLPs, RNNs, and CNNs on the Phishing Corpus and SpamAssassin datasets, achieving accuracy rates as high as 99.51\% accuracy and F1 scores as high as 99.52\%. 
Alshingiti et al.~\cite{alshingiti2023deep} explored multiple architectures, including LSTM-CNN hybrids on the Malicious Dataset 2016 \cite{mamun2016iscxurl2016}, achieving 97.6\% accuracy and an equal F1 score. Similarly, Muralidharan and Nissim \cite{muralidharan2023improving} combined CNN and BERT models using full email content from the VirusTotal \cite{virustotal} platform, reporting an AUC of 0.993 and a true positive rate (TPR) of 94.73\%.

In terms of model explainability, various techniques have been used to interpret deep learning models \cite{10849475}. In phishing detection, Uddin and Sarker \cite{uddin2024explainable} addressed the need for model transparency by introducing an explainable Transformer-based classifier built on DistilBERT, augmented with LIME \cite{ribeiro2016whyitrustyou} and Transformer Interpret to provide token-level explanations. Going further into the text domain, Grad-CAM has been adapted to generate visual explanations for deep learning models applied on NLP tasks. Selvaraju et al.~\cite{selvaraju2017grad} first proposed Grad-CAM to localize model attention in computer vision tasks. Subsequent research has explored its use in textual analysis, including legal text classification \cite{jain2019attention}, where visual attribution maps help interpret complex language inputs. These developments suggest that Grad-CAM and similar methods hold potential for interpreting HTML or email body content in phishing detection systems as well. However, to the best of our knowledge, Grad-CAM has not been applied to character-level models for phishing detection.

A limitation of machine learning models is their vulnerability to adversarial attacks \cite{ahmed2024comprehensivereviewadversarialattacks,10849376}. Gholampour and Verma \cite{10.1145/3579987.3586567} evaluate  the robustness of various phishing email detection models against adversarial attacks, such as Textfooler \cite{jin2019bert}, PWWS \cite{Waghela_2025}, DeepWordBug \cite{gao2018black}, and BAE \cite{Garg_2020}. 
Even transformer-based architectures like ALBERT \cite{lan2020albertlitebertselfsupervised}, RoBERTa \cite{liu2019robertarobustlyoptimizedbert} and BERT \cite{devlin2019bertpretrainingdeepbidirectional} are susceptible to such attacks, in some cases leading to misclassification rates that exceed 70\%. Abdullah and Khalifa~\cite{abdullah2024textemailspam} also studied the impact of adversarial attacks on spam and phishing email detection using CNN and LSTM models. They found that text-based manipulations, such as word substitutions and character perturbations, such as DeepWordBug, reduced model accuracy by up to 40\%.  Panum et al.~\cite{256938} analyzed how adversarial examples can bypass spam detection systems by simply modifying email content, for example, changing ``login'' to ``log1n''. Their experiments showed that traditional ML spam filters could be fooled in over 60\% of cases. They proposed adversarial training as a defense, which reduced the attack success rate to below 20\%. This line of work inspired us to test models under adversarial attacks and implement adversarial training to improve their robustness to such attacks.

\section{Methodology}

\subsection{Neural Phishing Detectors}
\vspace{-0.05cm}

As phishing detectors, we select three deep learning architectures: CharCNN, CharGRU, and CharBiLSTM. The neural models are adapted to operate at the character level instead of the word level. With this adaption, the models are able to capture textual patterns such as unusual punctuation, misspellings, and obfuscation techniques that are frequently used by attackers to bypass traditional filters. As suggested by Ionescu et al.~\cite{ionescu2014can}, character-level frameworks can often generalize better than word-level representations, as they can capture a broader variety of features.

For all three models (CharCNN, CharGRU, and CharBiLSTM), the character quantization process is identical, following the approach of Zhang et al.~\cite{zhang2015character}. Each email is first processed at the character level using a fixed alphabet consisting of 95 characters, including lowercase and uppercase English letters, digits, and special symbols. Each character is mapped to a unique index based on the character dictionary, and emails are converted into fixed-length input vectors with a maximum length of $T=1500$ characters. 
Shorter sequences are padded with zeros, while longer ones are truncated. The final input to the models consists of sequences of integer indices representing individual characters. These sequences are passed through an embedding layer, which maps each discrete index to a continuous trainable vector in a fixed-dimensional space. This transformation results in a dense matrix of shape $T \times D$, where $T$ is the sequence length, and $D$ is the embedding dimension. In this matrix, each row represents the learned embedding vector of a character. This embedded representation serves as the input to the respective deep learning models, enabling them to learn sequential or spatial patterns over characters. All models are equipped with a softmax layer. The binary class labels are transformed into one-hot encoding vectors and the training is performed via the cross-entropy loss. 

We further present the configuration for each architecture in more detail. We emphasize that the specific architectures are designed to operate in low-resource computational setups, e.g.~in a browser extension that may run locally, on a laptop with no GPU.


\subsubsection{CharCNN}
The CharCNN model comprises three convolutional layers, each with $64$ filters. Each layer is activated by thresholded Rectified Linear Units (ReLU). To improve generalization capacity, we employ dropout with a drop rate of $0.3$. 
We enhance the original CharCNN architecture proposed by Zhang et al.~\cite{zhang2015character} by integrating an attention mechanism based on Squeeze-and-Excitation (SE) \cite{hu2019squeezeandexcitationnetworks,butnaru2019morocomoldavianromaniandialectal, rogoz2021sarocodetectingsatirenovel}. The SE mechanism operates in three stages: squeeze, excitation, and scale. First, global average pooling is applied across the temporal dimension of each convolutional feature map, reducing each channel to a single scalar value that summarizes its global information (squeeze). This vector is then passed through a two-layer fully connected bottleneck: the first layer reduces dimensionality by a factor of $r$ (the reduction ratio), and the second restores it to the original number of channels. The use of ReLU and sigmoid activations allows the model to learn normalized channel-wise importance weights between 0 and 1 (excitation). Finally, these weights are used to rescale the original feature maps through element-wise multiplication (scale), allowing the model to emphasize the most informative channels, while suppressing less relevant ones. The SE mechanism enhances the capacity of the model to focus on specific character-level features used in phishing detection.


In our implementation, the SE block is conditionally applied at the end of specific convolutional layer groups, based on the configuration provided for each layer. Each convolutional layer is parameterized by five values: the number of filters, kernel size, optional pooling sizes, and an SE ratio. 


\subsubsection{CharGRU}
The CharGRU model belongs to the family of recurrent neural networks. Unlike the CharCNN model, which applies fixed-size convolutional filters to learn local patterns in the character space, the CharGRU architecture operates over time steps and is optimized to capture temporal dependencies across the full input sequence. This helps the model to identify dependencies, structural regularities, and cumulative context that emerge across character attributes particularly relevant to phishing detection, where attackers often use sequence-based obfuscation and social engineering patterns. 
The model has a Gated Recurrent Unit (GRU) layer with hidden units that manage information flow via update and reset gates. These gates allow the network to adaptively retain or discard past information. Character embeddings serve as input to the GRU layer, which outputs a sequence of hidden states. These are aggregated via global average pooling to produce a fixed-dimensional vector representation for the classification layer. 


\subsubsection{CharBiLSTM}
The CharBiLSTM architecture is another representative member of the RNN family, which is well-suited for modeling long-range sequential dependencies. The CharBiLSTM model receives character-level input sequences embedded into continuous vector representations, identical to the input pipeline used in CharGRU and CharCNN. These embeddings are processed by a BiLSTM layer with a predefined number of hidden units, allowing the model to construct an internal state that evolves as the character sequence progresses. The bidirectional version processes the sequence in both directions, left to right and right to left, thus being able to learn bidirectional relations among input characters. 

\vspace{-0.1cm}
\subsection{Model Explainability}
\vspace{-0.05cm}

To interpret the predictions of our character-level phishing detection models, we implement a character-level adaptation of Grad-CAM \cite{selvaraju2017grad}. Grad-CAM is a post-hoc, gradient-based explainability method that highlights which parts of the input are most influential in the decisions taken by a model.

Let $y^c$ be the score (logit) for class $c$ predicted by the model, and let $A^k \in \mathbb{R}^{T}$ denote the $k$-th feature map from a selected intermediate layer, where $T$ is the temporal (sequence) length, i.e.~the number of characters. Grad-CAM computes the gradient of the class score with respect to each feature map, as follows:
\begin{equation}
\frac{\partial y^c}{\partial A^k_t}, \quad \text{for } t = 1, \ldots, T.
\label{eq:grad}
\end{equation}
The importance weight for each feature map $A^k$ is calculated by global average pooling over time, as defined below:
\begin{equation}
\alpha_k^c = \frac{1}{T} \sum_{t=1}^{T} \frac{\partial y^c}{\partial A^k_t}.
\label{eq:alpha}
\end{equation}
The resulting weights are then combined with the corresponding feature maps to produce the class-discriminative localization map $L^c(t)$, using the ReLU activation to preserve only the positively contributing features, as follows:
\begin{equation}
L^c(t) = \texttt{ReLU} \left( \sum_k \alpha_k^c A^k_t \right).
\label{eq:score}
\end{equation}
In Eq.~\eqref{eq:score}, $L^c(t)$ assigns a relevance score to each timestep $t$, indicating the contribution of each character position to the final prediction for class $c$. We apply Grad-CAM to both convolutional (CharCNN) and recurrent (CharGRU, CharBiLSTM) architectures by selecting appropriate intermediate layers, namely the final convolutional block for the CharCNN, and the output sequence from GRU/BiLSTM layers. 
The resulting relevance maps are visualized as character-level heatmaps for interpretability. For a much better visualization, we export the character-level relevance scores $L^c(t)$ produced by Grad-CAM into HTML files, where each character in the email body is mapped to a red heatmap, where higher intensity indicates greater influence on the prediction of each model. 

\vspace{-0.1cm}
\subsection{Adversarial Attacks \& Defense}
\vspace{-0.05cm}

To evaluate the robustness of our phishing detection models, we apply character-level adversarial attacks generated with the DeepWordBug \cite{gao2018black} algorithm. The objective is to craft perturbed phishing emails that evade detection by fooling the classifier into predicting the ``clean'' class, similar with what attackers do (e.g., use ``1nput'' instead of ``input'').
We create a black-box attack scenario, where the attacker has no access to the internal weights of the model, but can query the model and observe the predicted class. The goal is to minimally modify phishing samples such that their classification is flipped from phishing to clean.
Given an input text sequence $x = [x_1, x_2, \ldots, x_T]$, we define a perturbation function $\delta(x)$, such that the modified input $x' = \delta(x)$ satisfies:
\begin{equation}
\arg\max f(x') \ne \arg\max f(x), \quad \text{and} \quad d(x, x') \leq \epsilon,
\label{eq:perturbation_condition}
\end{equation}
where $f(x)$ is the output of the model $f$ for input $x$, and $d(x, x')$ is the edit distance function (measured at the character level). We constrain $d(x, x')$ by perturbing no more than a fixed percentage $\epsilon$ of characters. In our case, $\epsilon = 10\%$ for testing data and $\epsilon = 20\%$ for adversarial training.
Following Gao et al.~\cite{gao2018black}, we use four types of character-level perturbations: swap, substitution, deletion, and insertion. 
For a sequence of length $T$, the number of modified characters is calculated as follows:
\begin{equation}
n = \lceil \epsilon \cdot T \rceil.
\label{eq:num_perturb}
\end{equation}

\begin{table}[t]
\centering
\caption{Large-scale dataset for phishing detection.}
\setlength\tabcolsep{0.5em}
\vspace{-0.2cm}
\begin{tabular}{|l|c|c|c|}
\hline
\textbf{Dataset} & \makecell{\textbf{Phishing} \\ \textbf{Emails}} & \makecell{\textbf{Clean} \\ \textbf{Emails}} & \makecell{\textbf{Total} \\ \textbf{Emails}} \\
\hline
CEAS 2008 Live Spam Challenge~\cite{ceas2008} & 21,842 & 17,312 & 39,154 \\
Enron Email Dataset~\cite{enron} & 13,976 & 15,791 & 29,767 \\
Ling-Spam Dataset~\cite{lingspam} & 458 & 2,401 & 2,859 \\
Nazario Phishing Email Dataset~\cite{nazario} & 1,565 & 0 & 1,565 \\
SpamAssassin Email Dataset~\cite{spamassassin} & 1,718 & 4,091 & 5,809 \\
TREC 2005 Public Spam Corpus~\cite{trec2005} & 22,932 & 32,278 & 54,210 \\
TREC 2006 Public Spam Corpus~\cite{trec2006} & 3,989 & 12,393 & 16,382 \\
TREC 2007 Public Spam Corpus~\cite{trec2007} & 29,392 & 24,353 & 53,745 \\
Nigerian 5 Phishing Email Dataset~\cite{nigerian5} & 1,500 & 0 & 1,500 \\
Nigerian Fraud Email Dataset~\cite{nigerianfraud} & 3,332 & 0 & 3,332 \\
Phishing Email Dataset~\cite{phishing} & 7,328 & 11,322 & 18,650 \\
\hline
Total & 108,032 & 119,941 & 227,973\\
\hline
\end{tabular}
\label{tab:email_dataset_statistics}
\vspace{-0.3cm}
\end{table}

\subsubsection{Adversarial Training}

To make our models more resistant to attacks, we also perform adversarial training by augmenting the training data with adversarial attacks. This involves taking 40\% of the phishing emails in the training set and applying the same character-level perturbations. These altered emails are added back into the training data, alongside the original samples. This augmentation gives the models a chance to learn from examples that mimic real adversarial scenarios. After training on the mixed dataset, we evaluate the models again on both clean and adversarial test data to assess their robustness against such attacks.

\section{Experiments and Results}
\label{sec:experimental_setup}

\subsection{Dataset}
\vspace{-0.05cm}

The large-scale dataset used in this study is obtained from multiple sources, as shown in Table~\ref{tab:email_dataset_statistics}. We collect a total of 227,973 emails: 108,032 phishing and 119,941 clean. We aggregate all data samples into a single database and remove duplicate emails and malformed entries. The final dataset includes business, academic and personal emails, offering a diverse representation of email types. Business emails come from the Enron corpus (emails of 150 senior managers of Enron \cite{enron}). Personal and general emails come from CEAS \cite{ceas2008}, Nazario (mass-market) \cite{nazario}, and Nigerian datasets (financial fraud) \cite{nigerian5, nigerianfraud,eze2024analysispreventionaibasedphishing}. Ling-Spam is an academic corpus composed of emails from a mailing list on linguistics, containing both legitimate academic messages and spam \cite{article20}.The remaining datasets do not explicitly specify whether the emails are personal, business, or otherwise.
We split the data into three subsets: 70\% for training, 15\% for validation, and 15\% for testing. This results in 162,938 emails for training, 34,916 for validation, and 34,915 for testing.

\vspace{-0.1cm}
\subsection{Computational Environment}
\vspace{-0.05cm}

The experiments are carried out under a clear computational constraint: all experiments are exclusively executed on CPU. More specifically, we use an Intel Core i7 with 16GB of RAM and an Intel Pentium Gold with 20GB of RAM for all tasks: analysis phase, model training, Grad-CAM explainability, adversarial attack and defense training. One of the objectives of this work is to demonstrate that it is possible to achieve competitive performance on phishing detection tasks even without the computational advantages of a GPU. This enables the deployment in low-resource settings, e.g.~in a browser extension.

\begin{table}[t]
\centering
\caption{Hyperparameter configurations for CharCNN, CharGRU, and CharBiLSTM.}
\label{tab:experimental_setup}
\setlength\tabcolsep{0.3em}
\vspace{-0.2cm}
\begin{tabular}{|l|c|p{2.5cm}|c|c|}
\hline
\textbf{Model} & \makecell{\textbf{Embed.}\\\textbf{Size}} & \makecell{\textbf{Architecture}\\\textbf{Details}} & \textbf{Optimizer} & \makecell{\textbf{Time /}\\\textbf{Epoch}} \\
\hline
CharCNN & 128 & 3 Conv1D layers (filters: 64, kernels: 5/3/1), SE Blocks, Thresholded ReLU & Nadam & $\approx37$ mins \\
\hline
CharGRU & 128 & 1 GRU layer (64 units), Global Average Pooling & Adam & $\approx46$ mins \\
\hline
CharBiLSTM & 32 & 1 BiLSTM layer (32 units), Global Average Pooling & Adam & $\approx41$ mins \\
\hline
\end{tabular}
\vspace{-0.2cm}
\end{table}

\begin{table*}[t]
\centering
\caption{Phishing detection performance levels of CharCNN, CharGRU, and CharBiLSTM, under three evaluation scenarios. The best score for each metric is in bold.}
\label{tab:model_performance}
\vspace{-0.2cm}
\begin{tabular}{|l|l|l|c|cc|cc|cc|}
\hline
\multirow{2}{*}{\textbf{Model}} & \multirow{2}{*}{\textbf{Training}} & \multirow{2}{*}{\textbf{Testing}} & \multirow{2}{*}{\textbf{Accuracy}} & \multicolumn{2}{c|}{\textbf{Precision}} & \multicolumn{2}{c|}{\textbf{Recall}} & \multicolumn{2}{c|}{\textbf{F1}} \\
& & &  & Weighted & Macro & Weighted & Macro & Weighted & Macro \\
\hline
CharCNN & clean & clean & 0.93711 & 0.83005 & 0.83738 & 0.75776 & 0.74575 & 0.73990 & 0.73628 \\
CharCNN & clean & adversarial & 0.91339 & 0.83508 & 0.84428 & 0.77197 & 0.75664 & 0.75623 & 0.75164 \\
CharCNN & adversarial & adversarial & 0.95237 & 0.91154 & 0.91616 & 0.89509 & 0.88883 & 0.89333 & 0.89234 \\
\hline
CharGRU & clean & clean & 0.97692 & 0.97698 & 0.97659 & 0.97692 & 0.97714 & 0.97692 & 0.97684 \\

CharGRU & clean & adversarial & 0.96019 & 0.95599 & 0.95676 & 0.95561 & 0.95436 & 0.95555 & 0.95531 \\
CharGRU & adversarial & adversarial & \textbf{0.98554} & \textbf{0.98539} & \textbf{0.98511} & \textbf{0.98534} & \textbf{0.98555} & \textbf{0.98534} & \textbf{0.98531} \\
\hline
CharBiLSTM & clean & clean & 0.96623 & 0.96623 & 0.96619 & 0.96623 & 0.96604 & 0.96623 & 0.96611 \\
CharBiLSTM & clean & adversarial & 0.95979 & 0.95698 & 0.95714 & 0.95695 & 0.95647 & 0.95694 & 0.95677 \\
CharBiLSTM & adversarial & adversarial & 0.98070 & 0.97604 & 0.97498 & 0.97534 & 0.97630 & 0.97536 & 0.97531 \\
\hline
\end{tabular}
\vspace{-0.1cm}
\end{table*}

\begin{table*}[t]
\centering
\caption{Performance levels, inference times and number of weights for CharGRU (trained with adversarial examples) versus LLaMA 3.2 (based on in-context learning with adversarial samples), on a subset of 1000 adversarial emails.}
\label{tab:llama_vs_chargru}
\vspace{-0.2cm}
\begin{tabular}{|l|c|cc|cc|cc|c|c|}
\hline
\textbf{Model} & \textbf{Accuracy} & \multicolumn{2}{c|}{\textbf{Precision}} & \multicolumn{2}{c|}{\textbf{Recall}} & \multicolumn{2}{c|}{\textbf{F1}} & \textbf{Inference} & \textbf{Trainable} \\
 &  & Weighted & Macro & Weighted & Macro & Weighted & Macro & \textbf{Time / Sample (s)} & \textbf{Params} \\
\hline
CharGRU & 0.94900 & 0.94924 & 0.94582 & 0.94900 & 0.94833 & 0.94907 & 0.94703 & 0.009 & 49.7K \\
LLaMA 3.2 & 0.76777 & 0.83316 & 0.80062 & 0.76777 & 0.80037 & 0.76772 & 0.76777 & 7.560 & 3.2B \\
\hline
\end{tabular}
\vspace{-0.3cm}
\end{table*}

\vspace{-0.1cm}
\subsection{Hyperparameter Configuration}
\vspace{-0.05cm}

All models are trained using the same pre-processing pipeline, with the same alphabet of 95 symbols and input sequences capped at 1500 characters. While the overall training configuration (30 epochs, batch size of 64, learning rate of 0.001, batch size of 64, and categorical cross-entropy loss) is the same, other hyperparameters, such as the embedding size and the number of hidden units, are adjusted based on hardware limitations to maintain training time at around 45 minutes per epoch (see Table~\ref{tab:experimental_setup}). CharBiLSTM, in particular, requires special attention: an initial configuration with 128-dimensional embeddings and 64 LSTM units resulted in training times of over 7 hours per epoch, which proved impractical. 

\vspace{-0.1cm}
\subsection{Experimental Scenarios}
\vspace{-0.05cm}

The performance of the models is analyzed under three distinct scenarios:
\begin{enumerate}
    \item Standard training and testing, where models are trained and evaluated on the clean data summarized in Table \ref{tab:email_dataset_statistics}.
    \item Standard training with adversarial testing, where models are trained on clean data, but evaluated on adversarially perturbed phishing emails via DeepWordBug.
    \item Adversarial training and testing, where adversarial examples are introduced during the training phase (40\% of phishing samples in the training set), and evaluation is performed on adversarial test data.
    \item Comparison with LLaMA 3.2 based on in-context learning on 1000 test emails.
\end{enumerate}

After applying the perturbations, each resulting adversarial email $x'$ is re-evaluated using the same metrics during adversarial testing and adversarial training and testing, namely accuracy, precision, recall, and F1 score. On the one hand, we compare the results between standard testing and adversarial testing to observe the vulnerability of each model under attacks. 
On the other hand, the adversarial training and testing scenario allows us to evaluate the effectiveness of model adaptation to adversarial patterns during learning. 

To better understand how our character-level models compare to transformer architectures, we compare our best-performing model with LLaMA 3.2 \cite{grattafiori2024llama3herdmodels} on a subset of 1000 emails from our dataset. We use prompt-based inference via the Ollama \cite{ollama2025} framework to classify them as either phishing or clean. This comparison allows us to assess differences in performance and decision consistency between neural architectures and large language models (LLMs) in a zero-shot setting.

\vspace{-0.1cm}
\subsection{Quantitative Results}
\vspace{-0.05cm}

In Table~\ref{tab:model_performance}, we report the results for the first three evaluation scenarios. 
CharGRU obtains the highest results in all scenarios, having an accuracy of 0.97 and an F1 score close to 0.98 on the clean test. These numbers drop to around 0.95 when the model is exposed to adversarial attacks. With the integration of adversarial training, the model outperforms its original scores, reaching around 0.985 for all metrics. The recall of this final CharGRU model confirms its ability to catch most of the phishing emails.

The CharBiLSTM model is just behind the CharGRU model, typically obtaining scores that are roughly 1\% below CharGRU, in all evaluation scenarios. Notably, the CharBiLSTM also exhibits performance improvements when it is trained with adversarial data, which further confirms the effectiveness of this training setup.

Compared to the RNN models, the CharCNN model obtains lower overall scores in all three scenarios. The CharCNN shows noticeable gaps, especially in terms of recall and F1. Although the adversarial training procedure leads to significant performance gains, the CharCNN remains well below the CharGRU and CharBiLSTM models across all evaluation metrics. The lower recall values of CharCNN across clean and adversarial datasets suggest that it is more prone to false negatives than the GRU and BiLSTM models, failing to catch a larger number of phishing emails. 

The performance gap between CharCNN and the recurrent models (CharGRU and CharBiLSTM) can be attributed to differences in how each architecture processes sequential information. The CharCNN model applies local convolutions over fixed-size windows, being inherently limited in modeling long-range dependencies across the input sequence. This constraint makes it less effective in capturing the distributed, context-sensitive patterns typical in phishing emails, especially when adversarial character-level changes are introduced. In contrast, CharGRU and CharBiLSTM both operate on the full sequence of characters, maintaining internal memory and capturing temporal dependencies over the entire input. This allows them to better preserve semantic and syntactic structure, even when small perturbations occur. 

We next compare our CharGRU (based on adversarial training) against LLaMA 3.2 \cite{grattafiori2024llama3herdmodels} on a subset of 1000 adversarial emails. The results are summarized in Table~\ref{tab:llama_vs_chargru}. Our CharGRU model outperforms LLaMA 3.2 in terms of all performance metrics, demonstrating higher precision and recall. Remarkably, the CharGRU model completed the task in just 9 seconds, while LLaMA required nearly 2 hours using prompt-based inference. The inference times were measured on a machine with an 11th Gen Intel(R) Core(TM) i7-1165G7 CPU, 16 GB RAM and no GPU. This indicates that the CharGRU model is about $800\times$ faster than LLaMA 3.2, while clearly outperforming LLaMA 3.2 in terms of all metrics. These results demonstrate that CharGRU is a robust and lightweight model, which is ready for immediate deployment in low-resource scenarios.

\vspace{-0.1cm}
\subsection{Explainability Analysis}
\vspace{-0.05cm}

\begin{figure}[t]
    \centering
    \includegraphics[width=1.0\linewidth]{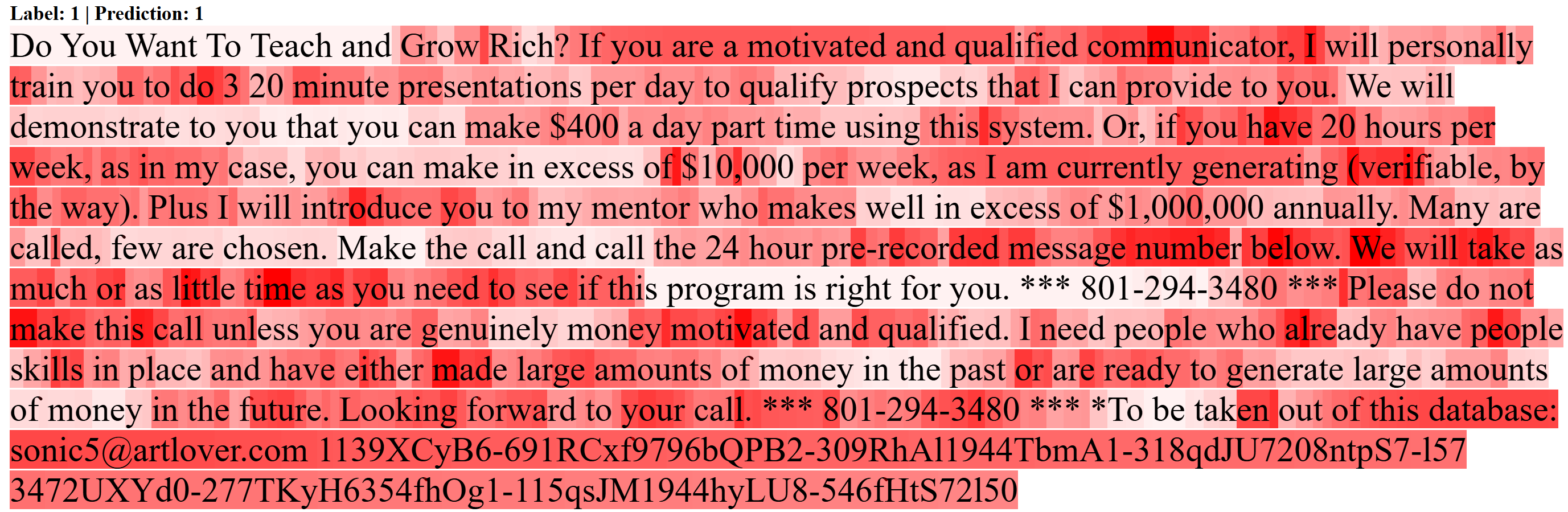} 
    \vspace{-0.7cm}
    \caption{Grad-CAM visualization of a true positive example.}
    \label{fig:tp1}
    \vspace{-0.2cm}
\end{figure}

\begin{figure}[t]
    \centering
    \includegraphics[width=1.0\linewidth]{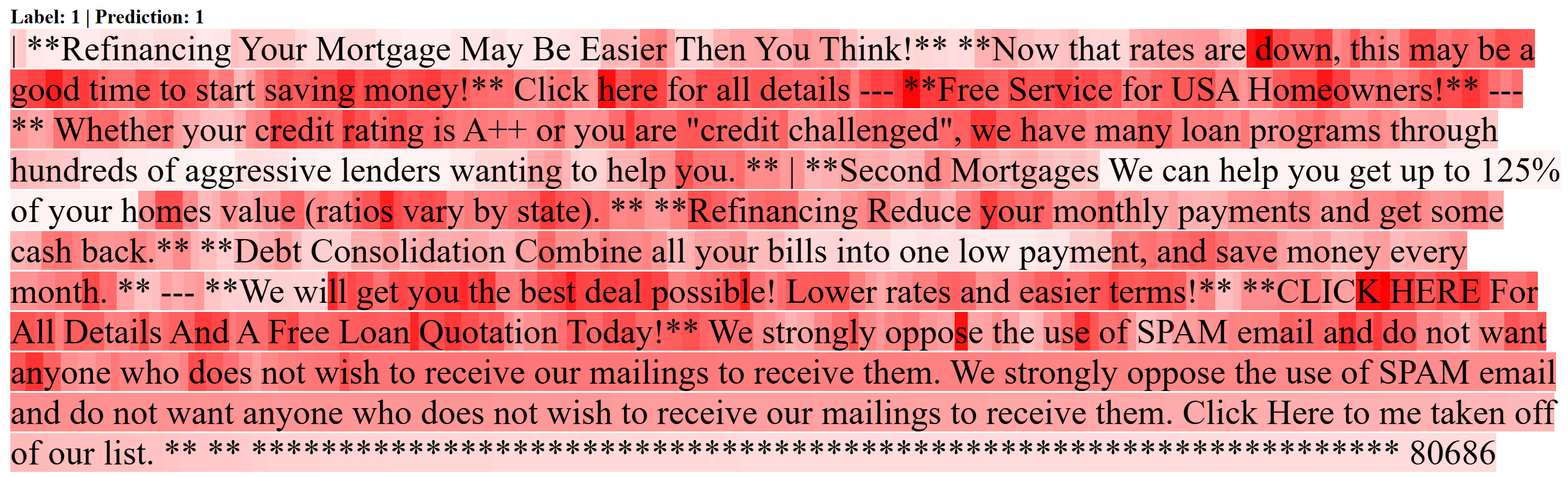} 
    \vspace{-0.7cm}
    \caption{Grad-CAM visualization of a true positive example.}
    \label{fig:tp2}
    \vspace{-0.2cm}
\end{figure}

\begin{figure}[t]
    \centering
    \includegraphics[width=1.0\linewidth]{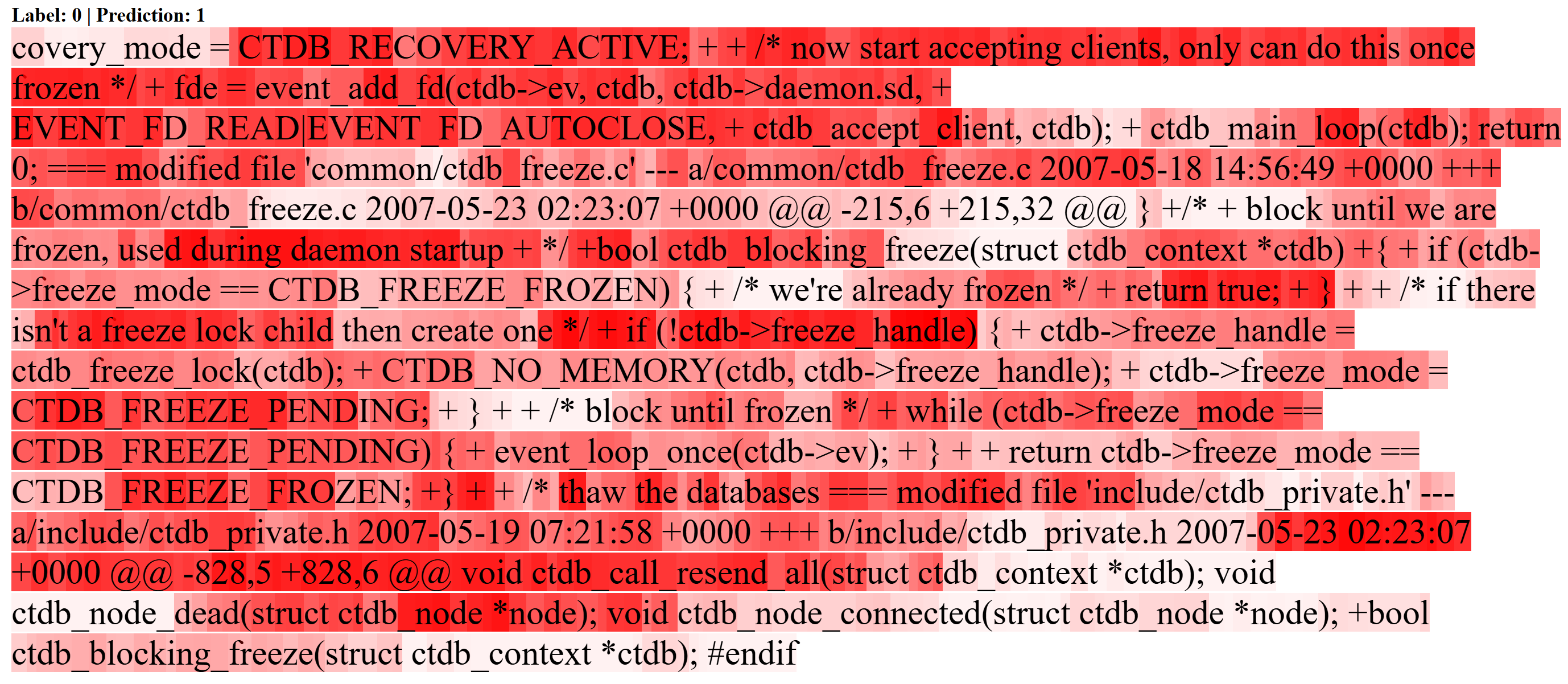} 
    \vspace{-0.7cm}
    \caption{Grad-CAM visualization of a false positive example.}
    \label{fig:fp1}
    \vspace{-0.2cm}
\end{figure}

We further visualize the Grad-CAM heatmaps on adversarial examples to investigate how perturbations alter the decisions of models. For the explainability analysis based on Grad-CAM, we generate HTML files to visualize the characters that influenced the decisions. As shown in Figures~\ref{fig:tp1} and~\ref{fig:tp2}, for true positive (phishing) samples, highly relevant regions include URLs, HTML tags, certain capitalized words (e.g. ``CLICK HERE''), and social engineering. In contrast, Figure~\ref{fig:fp1} illustrates a false positive case, where a legitimate email containing a code snippet was misclassified as phishing, due to the presence of numerous symbols, capitalized terms, and irregular patterns that mimic the visual structure of typical phishing content.

\section{Conclusion and Future Work}

In this study, we investigated character-level neural models for phishing email detection under standard and adversarial conditions. From a performance perspective, CharGRU demonstrated the most consistent and robust behavior, achieving the highest scores across all evaluation scenarios, especially after adversarial training. 
The adversarial testing showed that models can be vulnerable to character-level perturbations designed to evade detection. The performance drops observed during adversarial testing confirmed that even small textual modifications can mislead the models. After incorporating adversarial training, the models not only regained their original performance but also exceeded it, achieving their highest  scores in this configuration. This highlights the importance of incorporating adversarial mitigation techniques in phishing detection pipelines.

\section*{Acknowledgments}
This research is supported by the project ``Romanian Hub for Artificial Intelligence - HRIA'', Smart Growth, Digitization and Financial Instruments Program, 2021-2027, MySMIS no. 334906.

\bibliographystyle{IEEEtran}
\bibliography{bibliography} 
\end{document}